# New Insights into the Dynamics of Zwitterionic Micelles and Their Hydration Waters by Gigahertz-to-Terahertz Dielectric Spectroscopy

Deepu K. George,[1] Ali Charkhesht,[1] Olivia A. Hull,[2] Archana Mishra,[2] Daniel G. S. Capelluto,[3] Katie R. Mitchell-Koch,[2] Nguyen Q. Vinh[1*]

[1] Department of Physics and Center of Soft Matter and Biological Physics, Virginia Tech, Blacksburg, Virginia 24061

[2] Department of Chemistry, Wichita State University, Wichita, Kansas 67260

[3] Protein Signaling Domains Laboratory, Department of Biological Sciences, Biocomplexity Institute, and Center of Soft Matter and Biological Physics, Virginia Tech, Blacksburg, Virginia 24061

* corresponding author: Email: vinh@vt.edu; Phone: 540-231-3158

**Abstract:**

   Gigahertz-to-terahertz spectroscopy of macromolecules in aqueous environments provides an important approach for identifying their global and transient molecular structures, as well as directly assessing hydrogen-bonding. We report dielectric properties of zwitterionic dodecylphosphocholine (DPC) micelles in aqueous solutions over a wide frequency range, from 50 MHz to 1.12 THz. The dielectric relaxation spectra reveal different polarization mechanisms at the molecular level, reflecting the complexity of DPC micelle-water interactions. We have made a deconvolution of the spectra into different components and combined them with the effective-medium approximation to separate delicate processes of micelles in water. Our measurements demonstrate reorientational motion of the DPC surfactant head groups within the micelles, and two levels of hydration water shells, including tightly- and loosely-bound hydration water layers. From the dielectric strength of bulk water in DPC solutions, we found that the number of waters in hydration shells is approximately constant at $950 \pm 45$ water molecules per micelle in DPC concentrations up to 400 mM, and it decreases after that. At terahertz frequencies, employing the effective-medium approximation, we estimate that each DPC micelle is surrounded by a tightly-bound layer of $310 \pm 45$ water molecules that behave as if they are an integral part of the micelle. Combined with molecular dynamics simulations, we determine that tightly-bound waters are directly hydrogen-bonded to oxygens of DPC, while loosely-bound waters reside within 4 Å of micellar atoms. The dielectric response of DPC micelles at terahertz frequencies yields, for the first time, experimental information regarding the largest-scale, lowest frequency collective motions in micelles. DPC micelles are a relatively simple biologically relevant system, and this work paves the way for more insight in future studies of hydration and dynamics of biomolecular systems with gigahertz-to-terahertz spectroscopy.

## 1. INTRODUCTION

   Biological membrane components are amphipathic molecules, such as surfactants and lipids that can undergo surface and interfacial adsorption when dissolved in an aqueous solution. When surfactant molecules are dispersed in water, they aggregate to form micelles above a critical concentration, with the hydrophobic tails making up the core and hydrophilic head groups forming the shell. The separation of the hydrophobic and hydrophilic regions of micelles has been utilized extensively as an excellent tool to mimic biological environments and activities of lipid membranes.[1-3] Among a large variety of amphipathic molecules available for purifying and characterizing membrane proteins, the zwitterionic surfactant dodecylphosphocholine (DPC) (Fig. 1) forms spherical micelles with aggregation numbers of about $56 \pm 5$ at a concentration greater than the critical micelle concentration of 1 mM in aqueous solution.[4, 5] The zwitterionic surfactants are electrically neutral, but the charge they carry in the phosphocholine head group does influence the hydrophilic properties. Many zwitterionic surfactants have been used as model membrane systems because of characteristics such as the ability to form stable micelles,[6, 7] and the ability to bind to peptides and proteins while mimicking the anisotropic environment of a lipid membrane.[6-9] DPC micelles have a simpler structure[10] than most proteins, so studying the dynamics of micelles and their hydration layers offers an opportunity to determine processes that are not related to the specific structural vibrations of proteins. The outer micelle's hydrophilic surface is in contact with water, and thus, the system offers an opportunity to investigate the interaction of water with hydrophilic, biologically-relevant surfaces.

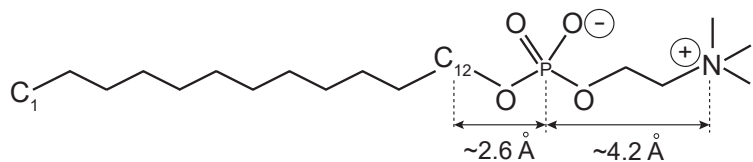

**Figure 1:** Chemical structure of DPC showing the numbering used in the text.

The dynamics of water in the hydration layer around proteins and other biomolecules play a crucial role in different aspects of biological processes. Some studies have concluded that water molecules in the hydration layer are rigidly attached to surfaces of molecules, resulting in an increase in the effective volume of the molecules.[11] On the other hand, evidence for the dynamic nature of the hydration layer is also abundant,[12] with some suggesting that there are fast and slow dynamic processes within the hydration layer. Some of the dynamical processes in proteins have been suggested as solvent slaved motions.[13] The significance of the hydration layers cannot be overstated in biological process and reactions, as they control the structure and function of biological systems.[14] There are many experimental techniques that allow the investigation of dynamics and structure of hydration water on a biomolecular surface. These include time-resolved fluorescence,[15] dielectric relaxation spectroscopy at gigahertz (GHz) frequencies,[16-19] nuclear magnetic resonance,[20] X-ray crystallography,[21] neutron scattering,[22] and infrared spectroscopy.[23] Among these techniques, dielectric spectroscopy from GHz to terahertz (THz) frequencies and computational techniques are advantageous for investigating the dynamics of water in confined systems, including interfacial or restricted environments, providing information on the hydrogen bonding, diffusion, and reorientation of water around DPC micelles as well as the dynamics of micelles themselves. The structural and dynamical properties of water in the relatively simple structure of DPC micelles will shed light on hydration dynamics in biological systems.

THz vibrational modes typically involve the low frequency, collective atomic motions of macromolecules, which include both inter- and intramolecular interactions. Thus, THz spectroscopy of biomolecules and lipid layers in aqueous environments provides an important approach for identifying their global and transient molecular structures as well as directly assessing hydrogen-bonding and other detailed environmental interactions.[24-26] However, a significant challenge in obtaining THz dielectric spectra of aqueous biomolecules and lipid layers is the strong absorption of water in the spectral range of 0.5 – 10 THz. The dielectric relaxation of surfactant micellar solutions have been reported,[16, 27, 28] but only below 89 GHz, hence dealing with the fluctuation of ion distribution and rotational motion of polar molecules. Advances in GHz to THz technology call for a more thorough study of the dielectric response of such simple systems. Such a study can act as an important step in understanding the behavior of more complex biomolecular systems in the GHz to THz frequency range. Micellar solutions have been extensively used in microwave and THz spectroscopy, especially for studies on nano-confined water[29] in the form of water dispersed in reverse micellar solutions. THz spectroscopy of protein-containing reverse micelles has also been investigated in the past as an alternative approach for probing collective vibrational motions in solution.[30] Recent developments in diode based frequency multipliers have improved the accuracy of THz measurements by several orders of magnitude, which allows for high-precision measurements of the strong absorption of aqueous solutions. In this paper, we investigate interactions of zwitterionic DPC micelles with water, using a spectrometer with a frequency range from 0.05 GHz to 1.12 THz.

## 2. MATERIAL AND METHODS

### 2.1. Materials and solution preparation

Dodecylphosphocholine (DPC), purchased from Anatrace (Cat. No. 29557), was used to prepare micellar solutions. DPC (m.w. = 351.5) was dissolved in deionized water to form micellar solutions with concentrations ranging from 50 to 800 mM. Accurate determination of the molarity of the solutions as well as the volume filling factor of DPC in solution was critical in our calculations. The solutions were prepared by weighing and measuring the volume after dissolving DPC in deionized water for several times to obtain an accurate value of the partial specific volume. The concentration of DPC in solutions was above the critical micelle concentration of 1 mM,[4, 5, 31] at which the liquid forms nearly spherical micelles with aggregation number of $56 \pm 5$.[4, 5] Results focus on the dynamics of water as well as the dynamics of DPC spherical micelles in solution. The chemical structure of DPC with the numbering used in the text is shown in the Fig. 1.

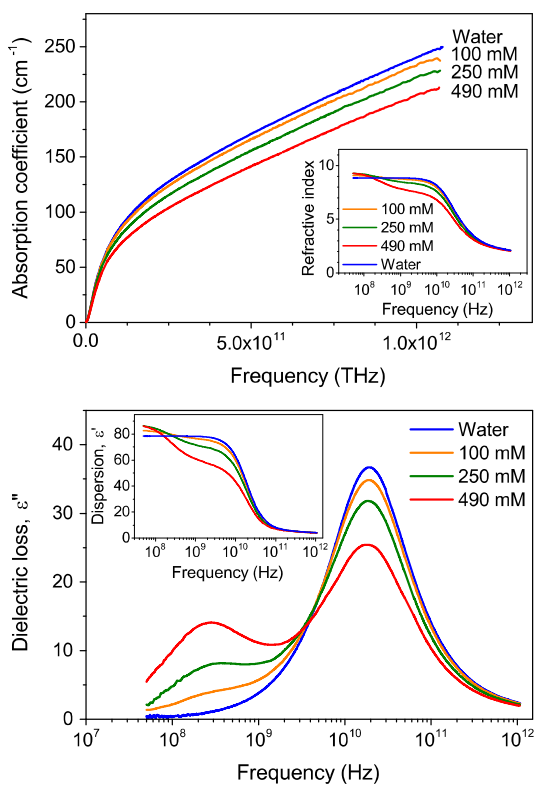

**Figure 2:** The interaction of DPC micelles with GHz to THz radiation provides insight into the liquid's dynamics over picosecond to nanosecond timescales. (**top**) The absorption spectra of both DPC micellar solutions and pure water increase monotonically with increasing frequency. The refractive indexes (**upper inset**) of DPC micelles and water, in contrast, decrease with increasing frequency. (**bottom**) The dielectric loss and the dielectric dispersion spectra (**lower inset**) from DPC aqueous solutions and pure water are obtained from absorption coefficient and refractive index measurements. Data were collected at 25°C.

### 2.2. Complex permittivity spectra

We have collected complex dielectric response spectra from DPC micellar solutions in a large range of frequencies, from 50 MHz to 1.12 THz, using two different methods. From 50 MHz to 50 GHz, a vector network analyzer (PNA N5225A) was combined with a dielectric probe (HP 85070E) and a transmission test set to measure the low frequency dielectric response (50 MHz – 5 GHz) of solutions. The cell of the transmission set consists of a coaxial line/circular cylindrical waveguide transition containing solutions. Air, water, and a shorting block were used as the standards for calibration of the dielectric probe. The real and imaginary parts of the dielectric constant were obtained directly from the system. The sample was kept in a sample cell made of anodized aluminum and the temperature was set at 25°C and controlled with an accuracy of ± 0.02°C using a Lakeshore 336 temperature controller.

The dielectric response of the samples at high frequency were studied using a GHz-to-THz spectrometer based on the above vector network analyzer together with frequency multipliers from Virginia Diodes, spanning the frequency range from 60 GHz to 1.12 THz. The setup is capable of simultaneously measuring intensity and phase over a large effective dynamical range of fifteen orders of magnitude.[32] Samples were kept in a home built variable path-length cell with submicron (~0.08 microns) precision in changing thickness. Temperature of the sample is controlled with the previously mentioned temperature controller. The sample cell was built with anodized aluminum to ensure thermal stability of solutions. For each frequency, we measured 200 data points of the intensity and phase shift of solutions as a function of the path-length. The absorption and refractive index of solutions at each frequency were determined from the best fit of the intensity and phase data to the sample thickness. The very high dynamic range of the frequency extenders, together with the precise sample thickness controller, allows us to obtain the most highly precise and accurate GHz-to-THz dielectric response spectra reported so far for this frequency range (Fig. 2).

With simultaneous measurements of the absorption and refractive index, the complex refractive index of a material can be expressed as

$$n^*(\nu) = n(\nu) + i\kappa(\nu) \qquad (1)$$

where $\nu$ is frequency, $n(\nu)$ is the refractive index of the solution, and $\kappa(\nu)$ the extinction coefficient of the solution. $\kappa(\nu)$ is related to the absorption coefficient, $\alpha(\nu)$, by $\kappa(\nu) = c\alpha(\nu)/(4\pi\nu)$ with $c$ being the speed of light. Similarly, the complex dielectric constant of a material can be expressed as

$$\varepsilon^*(\nu) = \varepsilon'(\nu) + i\varepsilon''(\nu) \qquad (2)$$

where $\varepsilon'(\nu)$ and $\varepsilon''(\nu)$ are the dielectric dispersion and dielectric loss components. Since our experiment can simultaneously measure both the absorption and refractive index of a material, $\varepsilon^*_{sol}(\nu)$, the complex dielectric response can be calculated from the following relations:

$$\begin{aligned}\varepsilon'_{sol}(\nu) &= n^2(\nu) - \kappa^2(\nu) = n^2(\nu) - (c\alpha(\nu)/4\pi\nu)^2 \\ \varepsilon''_{sol}(\nu) &= 2n(\nu) \cdot \kappa(\nu) = 2n(\nu)c\alpha(\nu)/4\pi\nu\end{aligned} \qquad (3)$$

From our absorption and refractive index measurement (Fig. 2, top), we have determined the dielectric spectra of DPC micellar solutions (Fig. 2, bottom).

**2.3. Molecular dynamics simulation details**

Molecular dynamics (MD) simulations have been performed using GROMACS package (version 4.5.3)[33] with cubic periodic boundary conditions. The force field for DPC used in this work was parametrized by Abel et al.[10] for GROMOS54A7 with SPC/E water model.[34] For initializing MD simulations, the topology file and coordinates of an equilibrated DPC micellar aggregate (comprised of 54 surfactant molecules) were acquired from Abel's website[35] and DPC force field parameters were used as published by Abel et al.[10]. First, the system was equilibrated for 50 ns at constant pressure (1 bar) and temperature (298 K) using Berendsen's barostat[36] and V-rescale thermostat.[37] Next, a 60.5 ns simulation was run at constant volume in a cube with length 7.44 nm, for data collection in a canonical ensemble using the Nosé-Hoover[38, 39] thermostat at T = 298 K. Water and surfactant were coupled separately, with a thermostat time constant of 0.4 ps. A time step of 2 fs was used to integrate the equations of motion. Bond lengths for the 54 DPC molecules were constrained using LINCS algorithm,[40] with SETTLE algorithm for the 12,794 water molecules.[41] Electrostatic interactions were calculated using the particle mesh Ewald method,[42] with an order of 4. Cutoffs of 1.4 nm were used for Coulombic and Lennard-Jones interactions. Analysis of MD data was carried out using GROMACS analysis tools and Octave.[43] The THz spectrum was calculated from MD simulations through a density of states (DoS) calculation (g_dos in the GROMACS package), which is acquired from the velocity autocorrelation function. The DoS analysis was run on the DPC surfactants in the micelle only (no waters contributing).

## 3. RESULTS AND DISCUSSION

### 3.1. Low frequency dielectric response (50 MHz to 50 GHz)

The dielectric response of polarization mechanisms in micellar solutions is still a matter of discussion, which requires more data collection on cationic, anionic, and zwitterionic surfactants. Complex dielectric responses of ionic surfactant micelles have been reported previously for cationic (hexadecyltrimethylammonium bromide - CTAB)[16] and anionic (sodium dodecyl sulfate – SDS)[28] micellar solutions (aqueous). In this paper, we have reported for the first time the dielectric response of zwitterionic DPC micellar solutions. The general observation for all three of these micelles is that a dielectric loss peak around 300 MHz (0.6 ns) is absent from the pure water spectrum, but present in micellar solutions. The amplitude of the dielectric loss increases with increasing micellar concentration. The responsible polarization mechanism at 0.6 ns was proposed to be due to the diffuse counterions.[16] It is noteworthy that the peak occurred essentially at the same frequency with similar shape and amplitude for both CTAB and SDS micelles. Later, the polarization mechanism was interpreted as the rotation of stable ion pairs[27, 44] or as hopping of counterions bound to the charged surface of the micelles.[45] However, given the similar observation for SDS, CTAB, and DPC, the fact that DPC micelles are zwitterionic, without counterions, makes this argument less likely. For DPC micelles, there will not be any electrical double layer formed by ionic clouds around micelles and hence the likelihood of a tangential counterion current, which seems to be a prerequisite for the dielectric response if we follow the argument in the above papers, is negligible. Thus, the results with the 0.6 ns timescale component in zwitterionic micelles indicate that counterions may not play an important role in the relaxation processes measured for cationic and anionic micelles. It is interesting to note the results of Tieleman *et al.*, who calculated the orientational correlation functions ($P_2$) of carbon atoms in the alkyl tails of surfactants in simulated DPC micelles, for comparison with NMR data.[46] They found timescales in the carbon rotation autocorrelation function on the order of hundreds of picoseconds. Since the environments of hydrophobic tails are similar within micelles, it is reasonable to expect that hydrophobic tails within different micelles may experience similar timescales of dynamics. It may also be reasonable to predict that different surfactants within micelles can experience similar reorientational (dipole moment) timescales. The reorientation of DPC surfactants within the micelle is discussed further below.

The dielectric properties of aqueous DPC micellar solutions at GHz frequencies show a complex behavior that originates from different polarization mechanisms at the molecular level (Fig. 2, bottom). The main loss peak frequency centered at ~19 GHz (~8.27 ps) remains virtually unchanged. This value was found to be the same as for pure water.[47-50] In this frequency range, the dielectric response of aqueous micellar solutions mainly has contributions from (i) the motion of macromolecules, *i.e.*, the rotation of micelles; (ii) the motion of surfactants within the micelle (*i.e.*, undulation and motion of head groups); (iii) hydration waters in the interfacial region surrounding DPC micelles (dipoles of water molecules in hydration shells); and (iv) the orientational polarization of bulk water molecules (water dipoles). The dielectric response of the orientational relaxation of molecular dipoles for macromolecules with the size of ~15 kDa, such as the entire DPC micelle, is typically in the range of 1 to 30 MHz.[51] We do not focus on these measurements in this report. The dielectric response of the orientational polarization of bulk water has been well established.[47, 49] However, the contribution of water in hydration shells of biomolecules is complicated. For example, the dielectric response of water in the hydration shells of lysozyme[51] and ribonuclease A[52] in aqueous solution consists of several dispersion regions. The contribution of water in hydration shells mainly originates from two kinds of hydration water, *i.e.*, tightly- and loosely-bound water. The first hydration layer consists of water strongly interacting with the macromolecular surface. The second layer, which has weaker interactions with the macromolecular surface or is not in direct contact, consists of loosely bound water molecules that exchange with the tightly-bound water and have dynamics approaching those of bulk water.

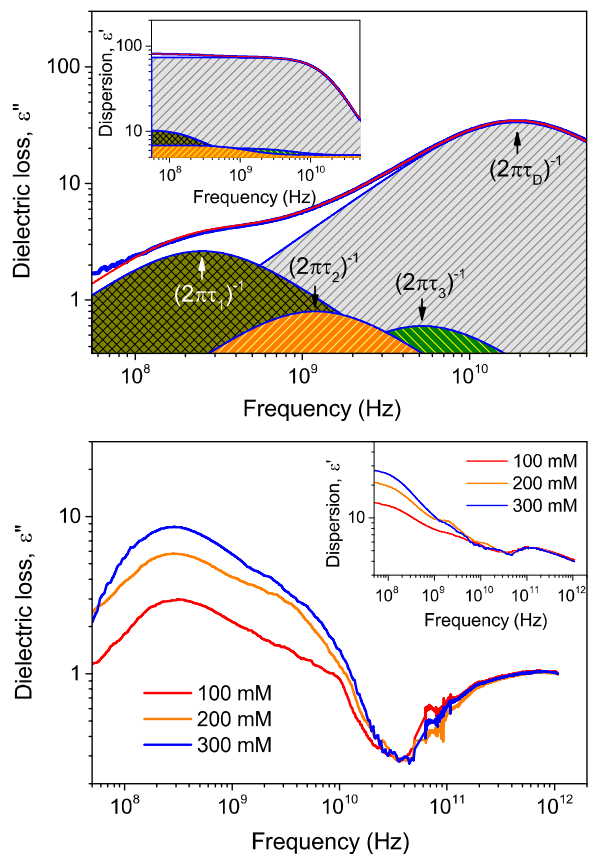

**Figure 3:** The dielectric loss and dielectric dispersion spectra of DPC aqueous solutions show relaxation processes at GHz frequencies. (**top**) The dielectric loss and dielectric dispersion (**upper inset**) spectra of 100 mM DPC in water provide insight into the dynamics of water molecules and micelles at the surface. The red curves are fits of the real and the imaginary components of the complex dielectric response. (**bottom**) The dielectric loss and dielectric dispersion spectra (**lower inset**) of the motion of surfactant head groups, the tightly- and loosely-bound water for several DPC micellar solutions have been obtained by subtracting the well-defined relaxation contribution of bulk water from the total spectrum. This procedure revealed their features in relaxation processes.

In the low frequency region, where librational motions and inertial effects do not contribute appreciably to the dielectric response, it is sufficient to consider Debye-type relaxations to analytically present our spectra within error limits. [Note that we have employed the Havriliak-Negami equations[53] to examine our data, but it did not show better results]. We have obtained dielectric response for the motion of surfactant head groups, and the tightly-bound, loosely-bound and bulk water in the form:

$$\varepsilon^*(\nu) = \varepsilon_\infty + \frac{\Delta\varepsilon_1}{1+i2\pi\nu\tau_1} + \frac{\Delta\varepsilon_2}{1+i2\pi\nu\tau_2} + \frac{\Delta\varepsilon_3}{1+i2\pi\nu\tau_3} + \frac{\Delta\varepsilon_D}{1+i2\pi\nu\tau_D} \qquad (4)$$

where $\varepsilon_0$ is the permittivity of free space. $\Delta\varepsilon_1$, $\Delta\varepsilon_2$, $\Delta\varepsilon_3$ and $\Delta\varepsilon_D$ are the dielectric strengths of each Debye process to the total relaxation for the motion of head groups on the micellar surfactant, the tightly-bound, the loosely bound, and the bulk water, respectively, while $\tau_1$, $\tau_2$, $\tau_3$ and $\tau_D$ are their relaxation times, respectively. $\varepsilon_\infty$ includes contributions to the dielectric response from modes at higher frequencies. The electrical conductivity of DPC micellar solutions is below the detection of our electric conductivity measurements. Thus, we do not include the *d.c.* electrical conductivity in the dielectric response in the Eq. 4. At 25°C, the well-known rotation relaxational time, $\tau_D$, for bulk water is typically 8.27 ps.[19, 47-50] The dielectric spectra of bulk water could be formally fitted by a superposition of two or three Debye processes,[47] yielding $\tau_D \sim$ 8.27 ps (19 GHz), $\tau_{D2} \sim$ 1.1 ps (145 GHz), and $\tau_{D3} \sim$ 0.178 ps (895 GHz) at 25°C. In this frequency range up to 50 GHz, the contributions of the dielectric response of bulk water at the high end of the frequency range are small, and can be neglected. The contributions from these modes at these high frequencies appear in the $\varepsilon_\infty$ parameter.

Using this method, we simultaneously fit the dielectric dispersion, $\varepsilon'$, and loss, $\varepsilon''$, with the same set of free parameters. Eight parameters in eq. 4 are varied simultaneously and the relaxation time for bulk water, $\tau_D$, is held fixed at the literature value.[47-50] Typical dielectric spectra for both dielectric dispersion, $\varepsilon'$, and dielectric loss, $\varepsilon''$, together with their spectral deconvolution are calculated for a DPC concentration of 100 mM (Fig. 3, top). The fit to the four-Debye model for GHz frequencies produces $\varepsilon_\infty = 5.1 \pm 0.5$, which is now within experimental uncertainty of the prior literature value[49, 50] and $\Delta\varepsilon_1$, $\Delta\varepsilon_2$, $\Delta\varepsilon_3$, $\Delta\varepsilon_D$ of $5.4 \pm 0.5$, $1.1 \pm 0.3$, $0.8 \pm 0.3$, and $68.6 \pm 0.3$, respectively. The dielectric loss spectrum obtained from the four-Debye model indicates four relaxation processes centered at $251 \pm 15$ MHz ($\tau_1 \sim 633 \pm 39$ ps), $1.38 \pm 0.13$ GHz ($\tau_2 \sim 105 \pm 10$ ps), $5.13 \pm 0.95$ GHz ($\tau_3 \sim 31 \pm 7$ ps), and $19.25 \pm 0.78$ GHz ($\tau_D \sim 8.27 \pm 0.35$ ps). The dominating $\tau_D$ process, the principle process for hydrogen-bonding liquids, reflects the cooperative reorientational dynamics of the dipole moment of bulk water in solution.

A more generalized model-independent approach for the low-frequency part of dielectric spectra has been achieved by subtracting the well-defined $\tau_D$ relaxation contribution from the total spectrum. The value for the dielectric strength, $\Delta\varepsilon_D$, and relaxation dynamics, $\tau_D$, obtained from the contribution of bulk water is then subtracted from the measured dielectric response, $\varepsilon^*(\nu)$. We have obtained dielectric response for several DPC micellar solutions (Fig. 3, bottom). This procedure revealed that dispersion and loss curves clearly exhibit their relaxation processes. The dielectric response at THz frequencies of DPC micellar solutions is complex, including the dielectric response of the collective motions of micelles and the librational and vibrational processes of water. As indicated below, we analyzed these dynamics in the high frequency response part separately, using the effective-medium approximation and MD simulations.

The accuracy in the evaluation of dielectric parameters, including dielectric strengths and their relaxation times, depends on the magnitude of the dielectric response, mainly on the micellar concentration. By fitting eq. 4 to experimental spectra, we have obtained dielectric parameters for the dielectric strengths of relaxation processes (Fig. 4, top), and their relaxation times, respectively (inset in Fig. 4, top), for several DPC micellar solutions. The time constants for relaxation processes are independent of micellar concentration. The amplitudes of the dielectric strengths vary strongly with rising concentration of micelles (Fig. 4). Specifically, while the dielectric strength for bulk water in micellar solutions at $\tau_D \sim 8.27$ ps decreases with increasing DPC concentration (Fig. 4, bottom), the dielectric strengths for slower processes increase with concentration. Further, the amplitudes for two processes, $\tau_2$ and $\tau_3$, show saturation at high micellar concentrations.

The dielectric measurements provide us information about relaxation processes in DPC micellar solutions. The relaxation process at the slowest time constant cannot be attributed to the rotational process of DPC micelles because $\tau_1$ is much smaller than the predicted relaxation timescale of macromolecules with a size of 15 kDa.[19, 51] Regarding whether this signal ($633 \pm 39$ ps) can be attributed to water dynamics at the micellar interface, it is instructive to consider the chemical makeup of the hydrophilic head groups. The DPC surfactants are electrically neutral, and the head groups contain several heteroatoms. However, the surface of DPC micelles provides fewer hydrogen bonds when compared to a protein surface. We can expect that the relaxation times for water in hydration layers of DPC micelles are faster than those of proteins, which typically exhibit slowest relaxation times on the order from 300 to 500 ps.[51, 52] As shown in Fig. 4, the dielectric strength for the slowest process, $\tau_1$, increases linearly for the whole range of DPC micellar concentrations. Thus, this relaxation has to be assigned to the rotation of the DPC surfactant, primarily, motion of the head groups. This interpretation is well supported by our dynamics studies of DPC head groups in the micellar environment by MD simulations.

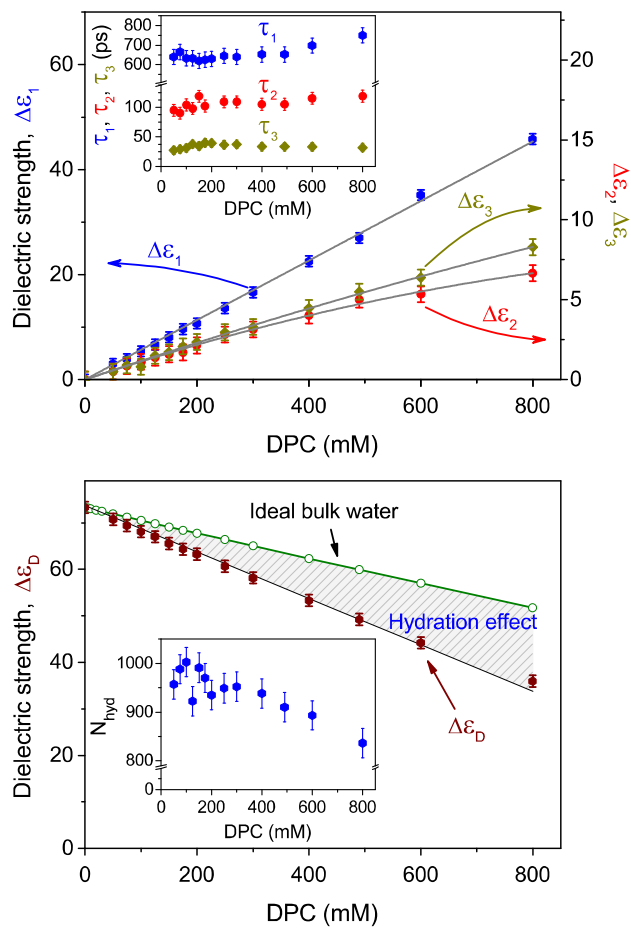

**Figure 4:** Waters' molecular-scale relaxations as a function of DPC micellar concentration, $c$, provides insight into their mechanistic relaxational processes. (**top**) The amplitudes of dielectric response of the motion of DPC head groups on the micellar surfactant, $\Delta\varepsilon_1$, tightly-bound water, $\Delta\varepsilon_2$, and loosely-bound water, $\Delta\varepsilon_3$, increase with rising DPC micellar concentration. The continuous lines serve as guides for the eye. The inset to the top shows their relaxation times, $\tau_1$, $\tau_2$ and $\tau_3$, respectively, as a function of DPC micellar concentration. (**bottom**) The dielectric strength of bulk water, $\Delta\varepsilon_D$, in DPC micellar solutions decreases with increasing DPC concentration. The continuous (green) line represents the ideal bulk-water dielectric amplitude from analysis of water concentration in solutions under an assumption that all water molecules in solution contribute to the bulk water process. The inset shows the hydration number as a function of DPC micelles concentration.

The dynamics of water in aqueous DPC micellar solutions present a complex dielectric response behavior at the molecular level. As mentioned above (Fig. 4), the amplitudes of the dielectric strengths for faster processes, $\tau_2$ and $\tau_3$, saturate at high micellar concentrations, thus these dynamics cannot be attributed to micelle-specific processes. The dynamics for these processes centered at $1 - 2$ GHz ($\sim 105$ ps) and $4 - 6$ GHz ($\sim 31$ ps), respectively, are greater than that of bulk water relaxation by factors of 13 and 4. From the behavior of dielectric strengths and the relaxation times, these processes are related to cooperative molecular dynamics in hydrated layers. The processes are highly cooperative in the densely packed hydration layers of DPC micelles, leading to a detectable slowdown in the dielectric relaxation times of bound water compared to that of bulk water. Water molecules having the relaxation time, $\tau_2$, of 105 ± 10 ps are in the tightly-bound hydration layer. These water molecules have a strong and direct hydration bond with DPC surfactant. The loosely bound water molecules, having indirect contacts or weaker hydration interactions with the macromolecular surface, have a relaxation time, $\tau_3$, of 31 ± 7 ps.

**TABLE 1:** Relaxation times, ($\tau_i$), and amplitudes, ($\Delta\varepsilon_i$), of dielectric response of the motion of head groups on the micellar surfactant, tightly-bound water, and loosely-bound water as well as the hydration number, N, per micelle.

| DPC (mM) | $\tau_1$ (ps) | $\tau_2$ (ps) | $\tau_3$ (ps) | $\Delta\varepsilon_1$ | $\Delta\varepsilon_2$ | $\Delta\varepsilon_3$ | $\Delta\varepsilon_D$ | N |
|---|---|---|---|---|---|---|---|---|
| 50 | 640 | 95 | 27 | 3.0 | 0.5 | 0.5 | 70.73 | 957 |
| 75 | 665 | 90 | 29 | 3.9 | 0.8 | 0.9 | 69.40 | 988 |
| 100 | 633 | 105 | 31 | 5.4 | 1.1 | 0.8 | 68.60 | 995 |
| 125 | 633 | 98 | 37 | 6.7 | 1.4 | 1.5 | 67.02 | 922 |
| 150 | 619 | 118 | 34 | 8.1 | 1.6 | 1.7 | 65.54 | 991 |
| 175 | 625 | 102 | 40 | 9.7 | 1.7 | 2.1 | 64.33 | 970 |
| 200 | 630 | 176 | 39 | 10.7 | 2.1 | 2.4 | 63.25 | 935 |
| 250 | 645 | 109 | 36 | 13.6 | 2.8 | 3.0 | 60.66 | 949 |
| 300 | 640 | 109 | 37 | 16.6 | 3.1 | 3.3 | 58.16 | 952 |
| 400 | 652 | 105 | 33 | 22.5 | 4.0 | 4.5 | 53.29 | 938 |
| 490 | 652 | 105 | 33 | 27.0 | 5.0 | 5.5 | 49.20 | 910 |
| 600 | 698 | 115 | 33 | 35.2 | 5.3 | 6.4 | 44.20 | 893 |
| 800 | 750 | 119 | 31 | 45.8 | 6.7 | 8.3 | 35.94 | 836 |

The structure of hydration shells, which are heterogeneous at the molecular level and distinct from bulk water, could be obtained from dielectric measurements. The hydration water, reflecting the total number of water molecules affected by macromolecules, can be deducted from the dielectric strength of bulk water. Therefore, this parameter is used to determine the water content in different kinds of soft materials.[51, 54] The presence of DPC micelles in aqueous solutions causes a decrease in the amplitude of the dielectric response of bulk water for two reasons: (i) The presence of the DPC micelles in solution reduces the volume of bulk water in solution, resulting in an overall lowering of the dielectric response. The continuous line (Fig. 4, bottom) represents the dielectric strength of ideal bulk water under an assumption that all water molecules in micellar solutions participate in the $\tau_D$ relaxation process of bulk water; (ii) Water molecules have hydrogen bonds in the vicinity of micelles and these water molecules no longer contribute to the $\tau_D$ relaxation process of bulk water. We estimated how many water molecules participated in hydration shells, through a comparison of the dielectric values for the bulk water in DPC solutions and the dielectric response of the total volume of water added to the solutions as a function of the concentration, $c$, of micellar solutions using the Cavell equation: [51, 52]

$$\Delta\varepsilon_j = \frac{\varepsilon}{\varepsilon+A_j(1-\varepsilon)} \frac{N_A c_j}{3k_B T \varepsilon_0} \frac{\mu_0^2}{(1-\alpha_j f_j)^2} \qquad (5)$$

The equation connects the dielectric strength of the $j$th relaxation process, $\Delta\varepsilon_j$, to the concentration of micelle in solutions, $c$, the shape parameter of the relaxing particle, $A_j$ (for sphere $A_j = 1/3$), the thermal energy, $k_B T$, Avogadro's number, $N_A$, the static permittivity, $\varepsilon$, the vacuum permittivity, $\varepsilon_0$, its permanent dipole, $\mu_0$, its polarizability, $\alpha_j$, and the reaction field factor, $f_j$. Normalizing eq. 5 to pure water, we obtain the apparent water concentration as a function of micellar concentration, $c_w^{app}(c)$:[19]

$$c_w^{app}(c) = \frac{\Delta\varepsilon_D(c)}{\Delta\varepsilon_{pure}} \frac{\varepsilon(0)(2\varepsilon(c)+1)(1-\alpha_w f_w(c))^2}{\varepsilon(c)(2\varepsilon(0)+1)(1-\alpha_w f_w(0))^2} c_{pure} \qquad (6)$$

where $\Delta\varepsilon_{pure} = 73.25$ is the dielectric strength of pure water at 25°C,[47-50] $c_{pure} = 55.35$ M represents the molarity of pure water. We define the effective hydration number per micelle, $N_{hyd}(c)$:

$$N_{hyd}(c) = \frac{c_w - c_w^{app}(c)}{n \cdot c} \qquad (7)$$

where $c_w$ is the concentration of water in the solution, and $n$ is the aggregation number of $56 \pm 5$ for DPC micelles. These water molecules per micelle do not participate in relaxation processes of bulk water due to the hydration effect (inset in Fig. 4, bottom). As can be seen, the hydration number is approximately constant at $950 \pm 45$ water molecules per micelle in the low range of DPC concentration up to 400 mM, but it decreases at higher DPC concentration. The relaxation parameters and hydration number per DPC micelle are summarized in Table 1.

When the DPC concentration is low, the micelles are monodisperse, and the average distance between micelles is much larger than the hydration number thickness, and thus, the hydration number, $N_{hyd}$, is, to a first approximation, independent of concentration. When the concentration of DPC increases to a certain level, the DPC micelles aggregate in small equilibrium clusters, resulting in a decrease of the hydration number. This observation is similar to dielectric measurements of lysozyme in a large range of concentrations.[51] Also, an observation using small-angle neutron scattering for DPC micelles has indicated decreasing micellar hydration with increasing concentration as surfactants become more closely packed and compete with one another for space within the micellar arrangement.[55]

### 3.2. High frequency response (60 GHz to 1.12 THz)

THz spectroscopy is a new tool to study solvation effects by probing the coupled collective modes of solute and solvent.[24, 47] It is experimentally challenging due to the strong THz absorption of water. Using our high resolution and dynamic THz frequency-domain spectroscopy and a variable-thickness cell, we have determined the absorption coefficient and refractive index of DPC micellar solutions along with that of pure water (Fig. 2). A quick glance at the absorption as well as the refractive index data indicates that these are strong functions of frequency, increasing and decreasing with increasing frequency, respectively. It is evident that the most prominent effect of addition of solvent is a monotonic decrease in absorption with increasing solvent concentration. This is primarily due to the fact that the higher absorbing solvent is replaced by the solute having a much lower absorption.

In dealing with the heterogeneous system, the dielectric response of dispersed solvent molecules in a bulk solvent is employed, rather than assuming that overall absorption is the sum of the absorption of its constituents. DPC micellar solutions are a mixture of water and DPC micelles and their complex dielectric response were determined from the experimental observables (Fig. 2, bottom). We assume that (i) DPC micelles in solution are spherical with a radius of $R_{micelles}$ and have a spherical hydration shell with a thickness of $d$; (ii) the water molecules in the hydration shell are a part of DPC micelles; (iii) solvent outside of the DPC micelles has the same dielectric property as that of pure water. Since the size of DPC micelles with hydration water is orders of magnitude smaller than the wavelength of the probing electromagnetic radiation, the medium can be considered homogenous with an effective dielectric response. A more elegant method has been employed for the effective-medium approximation, such as the Bruggeman model[56] which effectively treats both low and high concentration mixtures or Maxwell Garnet,[57] Wagner and Hanai approximations[58, 59] which are for low concentration limits. Following the Bruggeman approximation, the complex dielectric response of the solution can be determined from:

$$f_{micelles} \frac{\varepsilon^*_{micelles} - \varepsilon^*_{sol}}{\varepsilon^*_{micelles} + 2\varepsilon^*_{sol}} + (1 - f_{micelles}) \frac{\varepsilon^*_{wat} - \varepsilon^*_{sol}}{\varepsilon^*_{wat} + 2\varepsilon^*_{sol}} = 0 \qquad (8)$$

where $\varepsilon^*_{wat}(\nu)$ is the complex dielectric response of water; $\varepsilon^*_{micelles}$ is the complex dielectric response of hydrated DPC micelles described as the process of forming an aggregate with hydrophilic regions in contact with surrounding solvent; $f_{\text{micelles}} = (N_{\text{DPC}}/V)(4\pi/3)(R_{\text{micelles}} + d)^3$ is the volume fraction of the micelles with hydration water, and $N_{\text{DPC}}/V$ is the concentration of the DPC micelles in solution.

When we performed the Bruggemann effective-medium analysis, we found that each DPC micelle entraps a hydration shell composed of $310 \pm 45$ water molecules. Unlike the naïve absorbance-based method used in prior literature,[60] which estimate the size of the hydration shell by assuming that the macromolecule's absorption falls to zero at its minimum, this method of estimating the size of the tightly-bound hydration shell requires only the well-founded assumption that the micellar absorption falls to zero at zero frequency.[24] The value of 310 water molecules for the THz-defined hydration shell has similar size to a monolayer on the surface of the DPC micelles. Specifically, if we approximate DPC as a ~2 nm diameter sphere, a solvent layer with one molecule deep will contain 350 water molecules. Using the number of water molecules in the hydration shell related to the scaled filling factor and the measured $\varepsilon^*_{\text{wat}}(\nu)$ and $\varepsilon^*_{\text{sol}}(\nu)$, we employ eq. 8 to obtain the dielectric spectra for several hydrated DPC concentrations (Fig. 5). It should be noted that the hydration number calculated from the effective-medium approximation is lower than those from the GHz dielectric relaxation spectroscopy. This is expected, because through this method we determine the number of hydration water molecules that have strong hydrogen bonds with the DPC surfactant. These water molecules become an integral part of micelles and cannot move easily; instead, they are held in the tightly-bound hydration layer.

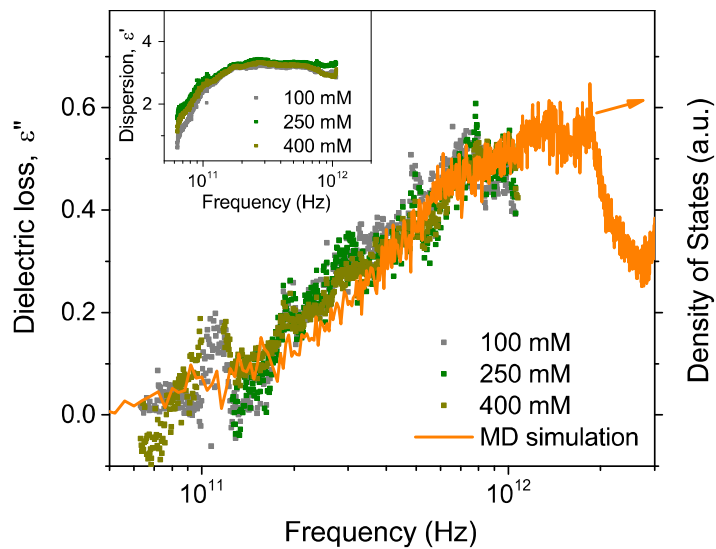

**Figure 5:** Dielectric loss, $\epsilon''(\nu)$, and dispersion, $\epsilon''(\nu)$, (**inset**) spectra of micelles in several DPC solutions at 25ºC in the THz frequency range from 60 GHz to 1.2 THz provide insight into the collective motions of micelles using the Bruggemann effective-medium approximation. From the effective-medium approximation, it is found that 310 water molecules in the hydration shell around DPC no longer behave as bulk water. The DoS analysis (orange line) from MD simulations was run on the DPC surfactants in the micelle only (no waters contributing).

Understanding how micellar dynamics and structure are connected to the chemical composition and geometry of the surfactants offers a considerable challenge. Numerous theoretical approaches and simulations[61-63] have been proposed to predict structure – property relationships. Experimental techniques typically obtained by small-angle neutral scattering,[64] static and dynamic light scattering[65], and cryogenic transmission electron microscopy,[66] which can probe a wide range of length- and time-scales, are needed to fully characterize micelles and correlate the micellar structure to the dynamical behavior, a fundamental prerequisite for developing practical formulations. The techniques measure the radius of gyration and make a link between micellar structure and dynamics. Here, we directly probe the collective dynamics of DPC micelles with THz radiation using the Bruggemann approximation to exclude the contribution from bulk water for several DPC solutions (Fig. 5). Upon doing so, we found that these spectra are characterized by a rising dielectric loss and a broad maximum of the dielectric dispersion component. For DPC concentrations below 400 mM, the measured dielectric response extracted from the effective-medium approximation is independent from concentration, suggesting that the size of the tightly-bound hydration shell is likewise independent of DPC concentration. This is the first time that the collective dynamics of DPC micelles at the THz region have been reported. MD simulations, discussed below, indicate that tightly-bound water molecules are those directly hydrogen-bonded to DPC surfactants. They also explain the collective motions of DPC micelles observed in the THz spectrum.

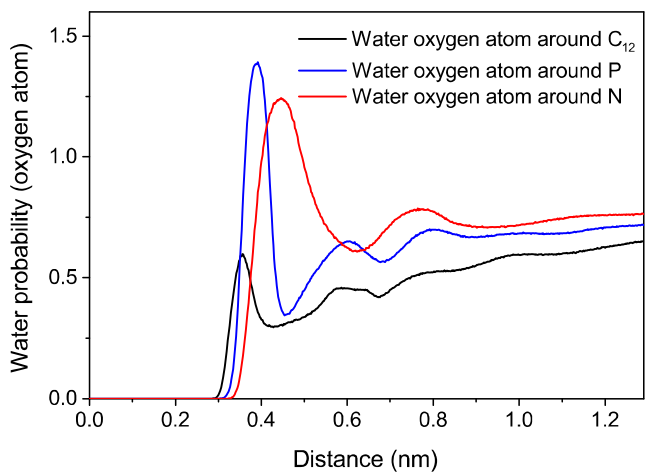

**Figure 6:** The solvent radial distribution functions (water oxygen atom) around C12 (black line), phosphorous (blue line), and nitrogen (red line) atoms of the DPC molecule.

### 3.3. Molecular dynamics simulations

To gain insight into the experimental observation of the dynamics and structure of hydration waters, the motion of head groups of DPC surfactants as well as the collective motions of micelles, we have carried out MD simulations for hydrated DPC micelles, for comparison to low concentration dielectric spectra. The combination of MD simulations with GHz to THz spectroscopy provides a microscopic picture of the coupled micelle-solvent dynamics associated with micellar aggregates. The solvation shells of DPC micelles were first analyzed by calculating the solvent radial density functions (Fig. 6). The radial distribution function (RDF) of water's oxygen was calculated around several atoms of DPC: $C_{12}$, the carbon nearest the head group; the phosphorous atom; and the nitrogen atom (see Fig. 1). The zwitterionic DPC head group has more chemical complexity than most surfactants, but the combined waters within the first peak of these three radial density functions appear to be the first solvation shell (comprised of both loosely-bound and tightly-bound waters, as discussed below). By taking a time average of the number of waters within the first peak of the solvent RDF around DPC head group atoms $C_{12}$, N, and P, the solvation shell size was calculated to be $995 \pm 26$ waters. This is in remarkably good agreement with the number of hydration waters ($950 \pm 45$) measured with dielectric spectroscopy at low concentrations of DPC micelle. We also found that by selecting waters within 4 Å of any DPC atom, the solvation shells have similar size, and thus, this definition was used for further analysis of solvation shell dynamics in DPC micellar simulations.

It was observed that the DPC lipid can have multiple head group conformations (Fig. 7).[62, 67] These include an extended monomer, in which the DPC molecule remains fairly linear (Fig. 7a); an intramolecular zwitterionic motif, in which the head group folds over by Coulombic attraction to itself (Fig. 7b); and a vicinal coupling motif, in which neighboring DPC molecules' cationic and anionic portions are in close proximity (Fig. 7c). Overall, the surfaces of DPC micelles are rough, having channels that are primarily lined with the cationic amine groups, with some anionic (phosphate) oxygens also being surface-exposed. The surface-exposed atoms are illustrated in Fig. 8, left panel, with trimethylamines pictured in aqua, and oxygens in red. Figure 8, right panel, shows waters within the channels, with the whole micelle surface pictured in dark blue. It can be seen that waters at the surface of the micelle are somewhat confined. Along with Coulombic interactions at the surface, this leads to a slowdown in water dynamics.

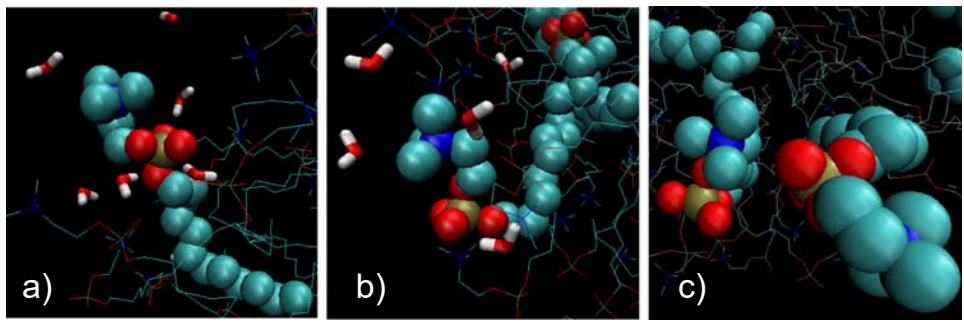

**Figure 7:** MD simulations show different conformational states of a DPC molecule. Solvation motifs: (a) extended monomer (b) intramolecular zwitterionic coupling (c) vicinal zwitterionic coupling.

Water molecules in the solvation layer around and within proteins have been found to have drastic reductions in dynamics, on the timescale of 500 ps by dielectric spectroscopy.[51] It is interesting to note that the cationic methylamine groups of DPC are much softer ions than are typically found in protein structures, with an effective radius larger than the $Rb^+$ cation. As discussed by Abel *et al.*, the methyl groups of the amine fully shield the nitrogen and, consequently, water molecules contact only the methyl groups on the cation.[10] Previous dielectric spectroscopy experiments have shown that larger monocations have a significantly smaller effect on water dynamics than harder, smaller ions such as $Na^+$ or $Li^+$.[52] Water around proteins also experiences the largest slowdown in dynamics within confined or concave spaces.[68] The divots or "canals" seen along the micelle surface are not as confining as protein active sites or interdomain regions. Furthermore, DPC only has hydrogen bond donors, rather than a multitude of hydrogen bond donors and acceptors, as is found in proteins. Therefore, it should not be surprising that no hydration dynamics around the timescale of 500 ps were found in the DPC micelle solvation shell. Rather, the longest dynamics measured in the GHz spectroscopy can be attributed to the motion of head groups of DPC surfactants within the micelle.

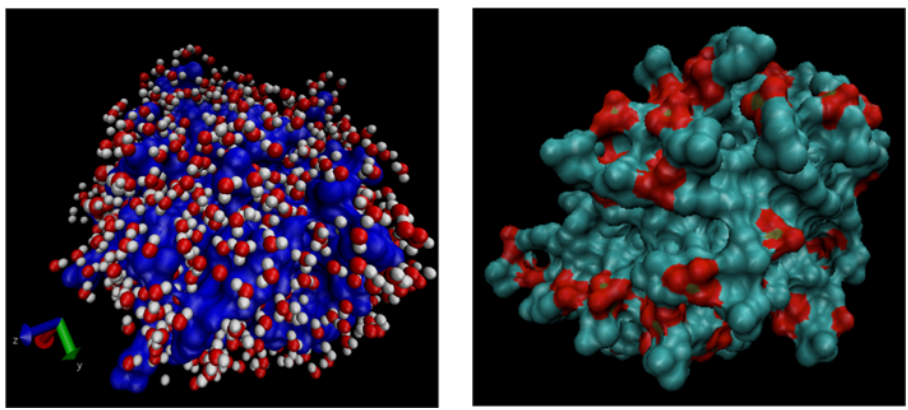

**Figure 8:** DPC micelle surface rendered (**left**) with alkyl groups (including trimethyl amine moieties) in aqua, oxygen in red, and phosphorous in gold (no waters are shown), (**right**) in dark blue, with solvation shell waters pictured in red.

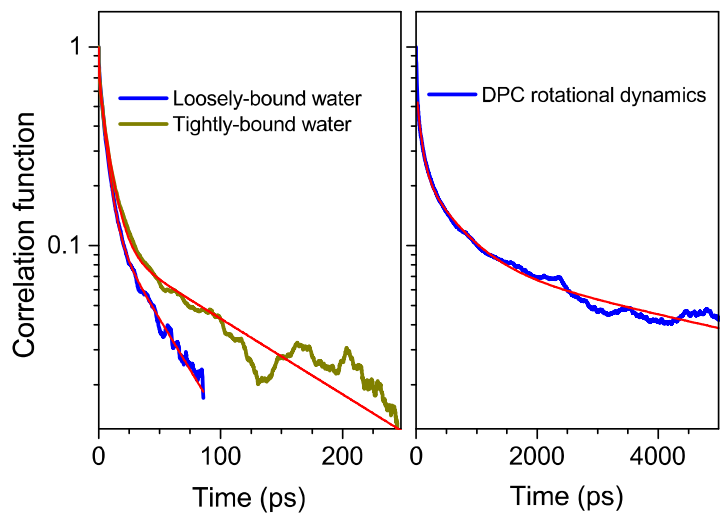

**Figure 9:** Rotational autocorrelation functions, $P_1(t)$ for hydration waters and DPC micelles show multiple-exponential decay behaviors. (**left**) The rotational autocorrelation functions of solvation shell waters hydrogen-bonded to DPC (dark yellow line) and other solvation shell waters (blue line) indicate a difference in the dynamics of tightly- and loosely-bound waters, respectively. (**right**) The rotational autocorrelation function of DPC monomers (blue line) within the micelle explains the dielectric response timescale from dynamics of DPC at 600 ps, arising primarily from the motion of surfactant head groups.

The only hydrogen bonding groups on the DPC molecule are the four oxygens around phosphorous, with the phosphate oxygens surrounded by a higher number of waters than the ester oxygens.[10] Waters with hydrogen bonds to surfaces typically result in slower reorientation times.[69] Therefore, solvation shell waters participating in hydrogen bonds with DPC oxygens were evaluated separately. It was found that, on average, there are 297 ± 17 hydrogen-bonded waters throughout the trajectory. This is in excellent agreement with the number of tightly-bound waters (310 ± 45) measured with dielectric spectroscopy at THz frequencies using the effective-medium approximation method.

The 1st Legendre Polynomial ($P_1$) of the reorientation autocorrelation function (ACF) for tightly-bound water (directly hydrogen bonded to DPC) and the reorientation of waters within the solvation shell that are not hydrogen bonded to DPC (identified as loosely bound waters) are given in Fig. 9a. According to the extended jump model, water reorientation is slower when it hydrogen bonds to a surface, and this is reflected in the slower decay of the rotational ACF of tightly-bound (hydrogen-bonded) waters.[70] The reorientational lifetimes of tightly- and loosely-bound waters were found by fitting a triexponential decay function. The long-time components of tightly- and loosely-bound waters were calculated to be 115 ± 5 and 42 ± 10 ps, respectively. These values are in good agreement with the dielectric spectroscopy measurements, which found characteristic timescales of 105 ps for tightly-bound waters and 31 ps for loosely-bound waters, respectively.

The relaxational dynamics of DPC molecules within the micelle were characterized with the rotational autocorrelation function (Fig. 9b). The reorientation timescales of head groups on DPC surfactants were determined by fitting a multi-exponential decay function (red traces, Fig. 9). As described in our recent paper on cetylpyridinium bromide micelles,[62] the short timescale of surfactant reorientation, fit to 79 ps for DPC, likely corresponds to spinning of the surfactant about its axis. Meanwhile, the longer timescale, fit at 591 ps, likely corresponds to head group reorientation or waving along its axis (like seaweed swaying in water). Note that the micelle environment confines surfactant rotation, approximately to a conical section, so that reorientation has a very long component of decay that fits to 4.9 ns. The THz-to-GHz spectroscopy of DPC micelles measures a component at 600 ps, which thus can be assigned to surfactant reorientation. It is interesting to recall that a 600 ps component has also been measured by dielectric spectroscopy at GHz frequencies for SDS[16] and CTAB[28] micelles. While those spectroscopic peaks have been attributed to the diffuse counterions,[16] the rotation of stable ion pairs,[27, 44] or the hopping of counterions bound to the charged surface of micelles,[45] it is also possible that the dipolar relaxation of surfactants of similar size, shape, and charge distribution within micelles takes place on a characteristic timescale on the order of ~600-1000 ps.

The contribution of the collective dynamics of DPC within the micellar environment to the THz spectrum can be seen from MD calculations for the density of states (orange curve, Fig. 5). The data from MD simulations come solely from the density of states calculation for the micelle itself (not including water). It can be seen that the collective dynamics of surfactant within the DPC micelle make major contributions to the THz spectrum. These motions within a micelle involve bending and undulation of surfactant molecules, including movements of the head group and curving of the alkyl tail.[62] These collective dynamics of micelles have been studied previously using neutron spin-echo spectroscopy[61] and light scattering,[71] with surfactants exhibiting breathing modes and worm-like motions.[71] The multiple conformations observed for the DPC head groups (Fig. 7) indicate that the dynamics of DPC within the micelle may be driven in part by interactions between the cationic and anionic portions of the head groups. The surfactants can transition between intramolecular and intermolecular (vicinal) zwitterionic coupling, as well as having extended conformations out into the hydration layer. In summary, these multiple conformations are evidence of the structural diversity and dynamic processes of surfactant within the DPC micelle. Future work in DPC simulations is planned to determine how water models influence head group conformational structure and dynamics.

## 4. CONCLUSIONS

We have conducted high-precision dielectric spectroscopy of DPC micellar solutions in a wide frequency range from 50 MHz to 1.12 THz to characterize the structure and dynamics of zwitterionic micelles and solvation, and used MD simulations to explain experimental results. The low frequency part indicates that four different relaxational processes occur in DPC aqueous solutions, including the motion of DPC surfactant head groups, which have a reorientation time of ~600 ps; hydration waters with times on the order of 30 to 150 ps; and bulk water with a time of 8 ps. There are two types of hydration water molecules in DPC micellar solutions, which are tightly- and loosely-bound waters at the DPC micellar surface. The total amount of hydration water of 950 molecules per micelle has been obtained from the dielectric strengths of bulk water in DPC solutions, which is in excellent agreement with the number of waters in MD simulations found within a distance of 4 Å from the DPC head group atoms. Using the effective-medium approximation for the dielectric measurements at THz frequency, we are able to extract the amount of tightly-bound waters in the first hydration shell of 310 molecules. The observation is in excellent agreement with MD simulations, which indicate ~300 waters directly hydrogen bonded to DPC. These water molecules have slower reorientational dynamics than the rest of the solvation shell.

It is interesting to note that, although the phosphatidylcholine head group is comprised of 11 heavy atoms (not H) and two ionic portions, the only hydration waters that are tightly bound are those in direct contact with the oxygen atoms (primarily, the two anionic oxygens). It is known that the dynamics of hydration waters can strongly influence the structure and dynamics of biomolecules, but our ability to predict how biomolecular structures influence water dynamics is quite limited. These results suggest that the confinement effects seen in proteins, which dramatically slow water dynamics, are not present in the rough terrain of the micellar surface. The cationic trimethylammonium groups likewise have a modest effect on water dynamics, being soft ions incapable of hydrogen bonding. Rather, the dominant structural factor affecting water dynamics in the presence of the DPC micelle is hydrogen bonding atoms. This is useful information for future work, when the effects of lipid membrane composition on water dynamics, and the interplay of hydration dynamics with lipid membrane and membrane protein properties, may be considered.

Finally, at the THz frequency, we have observed the collective vibrational modes within DPC micelles. Simulations indicate that the dominant contribution to the peaks comes from large-scale motions of the lipids within the confining environment of the micelle. These studies represent the first time that a clear peak has been experimentally observed and identified in the dielectric spectroscopy of membrane mimics.

**ACKNOWLEDGEMENT**

This work was supported by a grant from the Institute of Critical Technology and Applied Sciences (ICTAS) at Virginia Tech and the College of Liberal Arts and Sciences, Wichita State University. This project was supported by an Institutional Development Award (IDeA) from the National Institute of General Medical Sciences of the National Institutes of Health under grant number P20 GM103418. The content is solely the responsibility of the authors and does not necessarily represent the official views of the National Institute of General Medical Sciences or the National Institutes of Health.

**REFERENCES**

1. Deleu, M.; Crowet, J. M.; Nasir, M. N.; Lins, L., Complementary Biophysical Tools to Investigate Lipid Specificity in the Interaction between Bioactive Molecules and the Plasma Membrane: A Review. *Bba-Biomembranes* **2014,** *1838*, 3171-3190.
2. Eastoe, J.; Dalton, J. S., Dynamic Surface Tension and Adsorption Mechanisms of Surfactants at the Air-Water Interface. *Adv Colloid Interfac* **2000,** *85*, 103-144.
3. Bordag, N.; Keller, S., Alpha-Helical Transmembrane Peptides: A "Divide and Conquer" Approach to Membrane Proteins. *Chem Phys Lipids* **2010,** *163*, 1-26.
4. Lauterwein, J.; Bosch, C.; Brown, L. R.; Wuthrich, K., Physicochemical Studies of the Protein-Lipid Interactions in Melittin-Containing Micelles. *Biochim Biophys Acta* **1979,** *556*, 244-264.
5. Palladino, P.; Rossi, F.; Ragone, R., Effective Critical Micellar Concentration of a Zwitterionic Detergent: A Fluorimetric Study on n-Dodecyl Phosphocholine. *J Fluoresc* **2010,** *20*, 191-196.
6. Franzmann, M.; Otzen, D.; Wimmer, R., Quantitative Use of Paramagnetic Relaxation Enhancements for Determining Orientations and Insertion Depths of Peptides in Micelles. *Chembiochem* **2009,** *10*, 2339-2347.
7. Prive, G. G., Detergents for the Stabilization and Crystallization of Membrane Proteins. *Methods* **2007,** *41*, 388-397.
8. Warschawski, D. E.; Arnold, A. A.; Beaugrand, M.; Gravel, A.; Chartrand, E.; Marcotte, I., Choosing Membrane Mimetics for NMR Structural Studies of Transmembrane Proteins. *Bba-Biomembranes* **2011,** *1808*, 1957-1974.
9. Neumoin, A.; Arshava, B.; Becker, J.; Zerbe, O.; Naider, F., NMR Studies in Dodecylphosphocholine of a Fragment Containing the Seventh Transmembrane Helix of a G-Protein-Coupled Receptor from Saccharomyces Cerevisiae. *Biophys J* **2007,** *93*, 467-482.
10. Abel, S.; Dupradeau, F. Y.; Marchi, M., Molecular Dynamics Simulations of a Characteristic DPC Micelle in Water. *J Chem Theory Comput* **2012,** *8*, 4610-4623.
11. Bone, S.; Pethig, R., Dielectric Studies of Protein Hydration and Hydration-Induced Flexibility. *J Mol Biol* **1985,** *181*, 323-326.
12. Modig, K.; Liepinsh, E.; Otting, G.; Halle, B., Dynamics of Protein and Peptide Hydration. *J Am Chem Soc* **2004,** *126*, 102-114.
13. Fenimore, P. W.; Frauenfelder, H.; McMahon, B. H.; Parak, F. G., Slaving: Solvent Fluctuations Dominate Protein Dynamics and Functions. *P Natl Acad Sci USA* **2002,** *99*, 16047-16051.
14. Bellissent-Funel, M. C.; Hassanali, A.; Havenith, M.; Henchman, R.; Pohl, P.; Sterpone, F.; van der Spoel, D.; Xu, Y.; Garcia, A. E., Water Determines the Structure and Dynamics of Proteins. *Chem Rev* **2016,** *116*, 7673-7697.
15. Zhong, D. P.; Pal, S. K.; Zewail, A. H., Biological Water: A Critique. *Chem Phys Lett* **2011,** *503*, 1-11.
16. Barchini, R.; Pottel, R., Counterion Contribution to the Dielectric Spectrum of Aqueous-Solutions of Ionic Surfactant Micelles. *J Phys Chem-Us* **1994,** *98*, 7899-7905.

17. Buchner, R.; Baar, C.; Fernandez, P.; Schrodle, S.; Kunz, W., Dielectric Spectroscopy of Micelle Hydration and Dynamics in Aqueous Ionic Surfactant Solutions. *J Mol Liq* **2005,** *118*, 179-187.
18. Itatani, S.; Shikata, T., Dielectric Relaxation Behavior of Aqueous Dodecyldimethylamineoxide Solutions. *Langmuir* **2001,** *17*, 6841-6850.
19. Sato, T.; Sakai, H.; Sou, K.; Buchner, R.; Tsuchida, E., Poly(ethylene glycol)-Conjugated Phospholipids in Aqueous Micellar Solutions: Hydration, Static Structure, and Interparticle Interactions. *J Phys Chem B* **2007,** *111*, 1393-1401.
20. Kubinec, M. G.; Wemmer, D. E., NMR Evidence for DNA Bound Water in Solution. *J Am Chem Soc* **1992,** *114*, 8739-8740.
21. Ritland, H. N.; Kaesberg, P.; Beeman, W. W., An X-Ray Investigation of the Shapes and Hydrations of Several Protein Molecules in Solution. *J Chem Phys* **1950,** *18*, 1237-1242.
22. Svergun, D. I.; Richard, S.; Koch, M. H. J.; Sayers, Z.; Kuprin, S.; Zaccai, G., Protein Hydration in Solution: Experimental Observation by X-Ray and Neutron Scattering. *P Natl Acad Sci USA* **1998,** *95*, 2267-2272.
23. Holler, F.; Callis, J. B., Conformation of the Hydrocarbon Chains of Sodium Dodecyl-Sulfate Molecules in Micelles - an FTIR Study. *J Phys Chem-Us* **1989,** *93*, 2053-2058.
24. Vinh, N. Q.; Allen, S. J.; Plaxco, K. W., Dielectric Spectroscopy of Proteins as a Quantitative Experimental Test of Computational Models of Their Low-Frequency Harmonic Motions. *J Am Chem Soc* **2011,** *133*, 8942-8947.
25. Kindt, J. T.; Schmuttenmaer, C. A., Far-Infrared Dielectric Properties of Polar Liquids Probed by Femtosecond Terahertz Pulse Spectroscopy. *J Phys Chem-Us* **1996,** *100*, 10373-10379.
26. Markelz, A.; Whitmire, S.; Hillebrecht, J.; Birge, R., THz Time Domain Spectroscopy of Biomolecular Conformational Modes. *Phys Med Biol* **2002,** *47*, 3797-3805.
27. Shikata, T.; Imai, S., Dielectric Relaxation of Surfactant Micellar Solutions. *Langmuir* **1998,** *14*, 6804-6810.
28. Fernandez, P.; Schrodle, S.; Buchner, R.; Kunz, W., Micelle and Solvent Relaxation in Aqueous Sodium Dodecylsulfate Solutions. *Chemphyschem* **2003,** *4*, 1065-1072.
29. Boyd, J. E.; Briskman, A.; Sayes, C. M.; Mittleman, D.; Colvin, V., Terahertz Vibrational Modes of Inverse Micelles. *J Phys Chem B* **2002,** *106*, 6346-6353.
30. Murakami, H.; Toyota, Y.; Nishi, T.; Nashima, S., Terahertz Absorption Spectroscopy of Protein-Containing Reverse Micellar Solution. *Chem Phys Lett* **2012,** *519-20*, 105-109.
31. Kallick, D. A.; Tessmer, M. R.; Watts, C. R.; Li, C. Y., The Use of Dodecylphosphocholine Micelles in Solution NMR. *J Magn Reson Ser B* **1995,** *109*, 60-65.
32. George, D. K.; Charkhesht, A.; Vinh, N. Q., New Terahertz Dielectric Spectroscopy for the Study of Aqueous Solutions. *Rev Sci Instrum* **2015,** *86*, 123105-6.
33. Hess, B.; Kutzner, C.; van der Spoel, D.; Lindahl, E., GROMACS 4: Algorithms for Highly Efficient, Load-Balanced, and Scalable Molecular Simulation. *J Chem Theory Comput* **2008,** *4*, 435-447.
34. Berendsen, H. J. C.; Grigera, J. R.; Straatsma, T. P., The Missing Term in Effective Pair Potentials. *J Phys Chem-Us* **1987,** *91*, 6269-6271.
35. http://st-abel.com/downloads.htm.
36. Berendsen, H. J. C.; Postma, J. P. M.; Vangunsteren, W. F.; Dinola, A.; Haak, J. R., Molecular-Dynamics with Coupling to an External Bath. *J Chem Phys* **1984,** *81*, 3684-3690.
37. Bussi, G.; Donadio, D.; Parrinello, M., Canonical Sampling Through Velocity Rescaling. *J Chem Phys* **2007,** *126*, 014101-7.
38. Hoover, W. G., Canonical Dynamics - Equilibrium Phase-Space Distributions. *Phys Rev A* **1985,** *31*, 1695-1697.
39. Nose, S., A Molecular-Dynamics Method for Simulations in the Canonical Ensemble. *Mol Phys* **1984,** *52*, 255-268.
40. Hess, B., P-LINCS: A Parallel Linear Constraint Solver for Molecular Simulation. *J Chem Theory Comput* **2008,** *4*, 116-122.

41. Miyamoto, S.; Kollman, P. A., Settle - an Analytical Version of the Shake and Rattle Algorithm for Rigid Water Models. *J Comput Chem* **1992,** *13*, 952-962.
42. Essmann, U.; Perera, L.; Berkowitz, M. L.; Darden, T.; Lee, H.; Pedersen, L. G., A Smooth Particle Mesh Ewald Method. *J Chem Phys* **1995,** *103*, 8577-8593.
43. Eaton, J. W.; Bateman, D.; Hauberg, S.; Wehbring, R., *GNU Octave Version 3.8.1 Manual: a High-Level Interactive Language for Numerical Computations. Create Space Independent Publishing Platform. ISBN 1441413006, URL http://www.gnu.org/software/octave/doc/interpreter/* **2014**.
44. Imai, S.; Shiokawa, M.; Shikata, T., Dielectric Relaxation Behavior of Cationic Micellar Solutions: 2. *J Phys Chem B* **2001,** *105*, 4495-4502.
45. Baar, C.; Buchner, R.; Kunz, W., Dielectric Relaxation of Cationic Surfactants in Aqueous Solution. 2. Solute Relaxation. *J Phys Chem B* **2001,** *105*, 2914-2922.
46. Tieleman, D. P.; van der Spoel, D.; Berendsen, H. J. C., Molecular Dynamics Simulations of Dodecylphosphocholine Micelles at Three Different Aggregate Sizes: Micellar Structure and Chain Relaxation. *J Phys Chem B* **2000,** *104*, 6380-6388.
47. Vinh, N. Q.; Sherwin, M. S.; Allen, S. J.; George, D. K.; Rahmani, A. J.; Plaxco, K. W., High-Precision Gigahertz-to-Terahertz Spectroscopy of Aqueous Salt Solutions as a Probe of the Femtosecond-to-Picosecond Dynamics of Liquid Water. *J Chem Phys* **2015,** *142*, 164502-7.
48. Fukasawa, T.; Sato, T.; Watanabe, J.; Hama, Y.; Kunz, W.; Buchner, R., Relation Between Dielectric and Low-Frequency Raman Spectra of Hydrogen-Bond Liquids. *Phys Rev Lett* **2005,** *95*, 197802-4.
49. Ellison, W. J., Permittivity of Pure Water, at Standard Atmospheric Pressure, over the Frequency Range 0-25 THz and the Temperature Range 0-100 Degrees C. *J Phys Chem Ref Data* **2007,** *36*, 1-18.
50. Buchner, R.; Hefter, G. T.; May, P. M., Dielectric Relaxation of Aqueous NaCl Solutions. *J Phys Chem A* **1999,** *103*, 1-9.
51. Cametti, C.; Marchetti, S.; Gambi, C. M. C.; Onori, G., Dielectric Relaxation Spectroscopy of Lysozyme Aqueous Solutions: Analysis of the Delta-Dispersion and the Contribution of the Hydration Water. *J Phys Chem B* **2011,** *115*, 7144-7153.
52. Oleinikova, A.; Sasisanker, P.; Weingartner, H., What Can Really Be Learned from Dielectric Spectroscopy of Protein Solutions? A Case Study of Ribonuclease A. *J Phys Chem B* **2004,** *108*, 8467-8474.
53. Havrilia, S.; Negami, S., A Complex Plane Representation of Dielectric and Mechanical Relaxation Processes in Some Polymers. *Polymer* **1967,** *8*, 161-210.
54. Hayashi, Y.; Katsumoto, Y.; Omori, S.; Kishii, N.; Yasuda, A., Liquid Structure of the Urea-Water System Studied by Dielectric Spectroscopy. *J Phys Chem B* **2007,** *111*, 1076-80.
55. Pambou, E.; Crewe, J.; Yaseen, M.; Padia, F. N.; Rogers, S.; Wang, D.; Xu, H.; Lu, J. R., Structural Features of Micelles of Zwitterionic Dodecyl-Phosphocholine (C12PC) Surfactants Studied by Small-Angle Neutron Scattering. *Langmuir* **2015,** *31*, 9781-9789.
56. Bruggemann, D. A. G., Berechnung Verschiedener Physikalischer Konstanten von Heterogenen Substanzen. *Ann. Phys. Leipzig* **1935,** *24*, 636-664.
57. Garnett, J. C. M., Colours in Metal Glasses and in Metallic Films. *Philos T R Soc Lond* **1904,** *203*, 385-420.
58. Hanai, T., Theory of the Dielectric Dispersion due to the Interfacial Polarization and its Application to Emulsions *Colloid Polym Sci* **1960,** *171*, 23-31.
59. Choy, T. C., *Effective Medium Theory: Principle and Applications* **1999**.
60. Xu, J.; Plaxco, K. W.; Allen, S. J., Collective Dynamics of Lysozyme in Water: Terahertz Absorption Spectroscopy and Comparison with Theory. *J Phys Chem B* **2006,** *110*, 24255-24259.
61. Farago, B.; Monkenbusch, M.; Richter, D.; Huang, J. S.; Fetters, L. J.; Gast, A. P., Collective Dynamics of Tethered Chains - Breathing Modes. *Phys Rev Lett* **1993,** *71*, 1015-1018.
62. Verma, R.; Mishra, A.; Mitchell-Koch, K. R., Molecular Modeling of Cetylpyridinium Bromide, a Cationic Surfactant, in Solutions and Micelle. *J Chem Theory Comput* **2015,** *11*, 5415-5425.

63. Ahn, Y. N.; Mohan, G.; Kopelevich, D. I., Collective Degrees of Freedom Involved in Absorption and Desorption of Surfactant Molecules in Spherical Non-Ionic Micelles. *J Chem Phys* **2012,** *137*, 164902-18.
64. Jerke, G.; Pedersen, J. S.; Egelhaaf, S. U.; Schurtenberger, P., Flexibility of Charged and Uncharged Polymer-Like Micelles. *Langmuir* **1998,** *14*, 6013-6024.
65. Sato, T.; Einaga, Y., Dynamic Light Scattering from Non-Entangled Wormlike Micellar Solutions. *Langmuir* **2008,** *24*, 57-61.
66. Gonzalez, Y. I.; Kaler, E. W., Cryo-TEM Studies of Worm-Like Micellar Solutions. *Curr Opin Colloid In* **2005,** *10*, 256-260.
67. Wymore, T.; Gao, X. F.; Wong, T. C., Molecular Dynamics Simulation of the Structure and Dynamics of a Dodecylphosphocholine Micelle in Aqueous Solution. *J Mol Struct* **1999,** *485*, 195-210.
68. Hua, L.; Huang, X. H.; Zhou, R. H.; Berne, B. J., Dynamics of Water Confined in the Interdomain Region of a Multidomain Protein. *J Phys Chem B* **2006,** *110*, 3704-3711.
69. Sterpone, F.; Stirnemann, G.; Laage, D., Magnitude and Molecular Origin of Water Slowdown Next to a Protein. *J Am Chem Soc* **2012,** *134*, 4116-4119.
70. Fogarty, A. C.; Laage, D., Water Dynamics in Protein Hydration Shells: The Molecular Origins of the Dynamical Perturbation. *J Phys Chem B* **2014,** *118*, 7715-7729.
71. Cates, M. E.; Candau, S. J., Statics and Dynamics of Worm-Like Surfactant Micelles. *J Phys-Condens Mat* **1990,** *2*, 6869-6892.